# New interaction potentials for alkaline earth silicate and borate glasses

Yueh-Ting Shih[a], Siddharth Sundararaman[b], Simona Ispas[c] and Liping Huang[a,*]

[a]Department of Materials Science and Engineering, Rensselaer Polytechnic Institute, Troy, NY 12180, United States
[b]Lawrence Berkeley National Laboratory, Berkeley, CA 94720, United States
[c]Laboratoire Charles Coulomb (L2C), University of Montpellier, CNRS, Montpellier, France

**Abstract**

Structure and properties of magnesium silicate and borate melts and glasses were investigated by using newly parameterized interaction potentials in molecular dynamics simulations and compared with those of calcium silicate and borate. The competition between the depolymerization of the silica network and the formation of new bonds with modifier ions leads to the enhancement of the elastic moduli with increasing modifier content in alkaline earth silicate glasses. Compared with calcium silicate, the higher elastic moduli of magnesium silicate result from a higher connectivity of the overall glass network due to the incorporation of fourfold coordinated magnesium and a more rigid connection between $SiO_4$ tetrahedra and modifier ions. In contrast to the silicates, the effect of modifier content on the elastic moduli of alkaline earth borates is dominated by the fraction of fourfold coordinated boron ($N_4$). Calcium borate with higher $N_4$ shows a more rigid network structure and higher elastic moduli.

# 1. Introduction

Calcium silicates are commonly found in bioactive materials for medical treatment and cementitious materials for construction [1-5]. Previous studies indicated that substitution of CaO by MgO in silicate glasses modifies their chemical durability and increases the fracture toughness with a concomitant decrease of the Young's modulus [6-10]. The unique properties of magnesium-containing glass may be attributed to the distinctive role of magnesium in the glass network. Based on the classical glass formation theory [11], when added into the glass network, alkaline earth ions act as network modifiers to break the connectivity of the network and form non-bridging oxygens (NBOs). However, as magnesium is known to have a high field strength and a high electronegativity, its bond with oxygen exhibits some covalent characters and thus it behaves more like a network former, as shown in several experimental and simulation studies [8-10, 12-22]. From the infrared (IR) spectra of aluminosilicate glasses, the calculated ionicity is 0.9 for calcium but only 0.7 for magnesium, which explains the covalent character of magnesium-oxygen bonds [12]. The deconvolution of $^{29}$Si nuclear magnetic resonance (NMR) spectra of silicate-based bioactive glasses with the composition of $49.5SiO_2–1.1P_2O_5–23.0((1- x)CaO–xMgO)–26.4Na_2O$ suggests that, although 86% of the magnesium oxide acts traditionally as a network modifier, up to 14% of the magnesium oxide enters the glass network as $MgO_4$ tetrahedron, resulting in an increased polymerization of the glass network [9]. Similarly, previous molecular dynamics (MD) simulations of substitution of CaO with MgO in soda-lime silicate and silicate-based bioactive glasses showed that the higher field strength of Mg leads to a different structure compared to the effect of Ca [8, 10]. More specifically, Ca ions, coordinated by six oxygen atoms, act as network modifier, while some of the Mg ions are coordinated by four oxygen atoms to form $MgO_4$ tetrahedron and interconnect with the $SiO_4$ network.

For another important glass former, boron, the addition of MgO results in a unique structural change in the glass network as well. The fraction of fourfold coordinated boron ($N_4$) from NMR and FTIR results indicated a smaller amount of $N_4$ in magnesium borate compared with other alkaline earth borates at the same modifier content [23, 24]. Although, several studies investigated the structure and properties of the magnesium-containing silicate and borate, most of them focused on a narrow composition range [8, 10, 24]. The purpose of this work is to use MD simulation to systematically study the effect of MgO on structure and properties of silicate and borate melts and glasses in comparison with CaO. To this end, reliable interaction potentials are needed to model these systems over a large composition range and under different thermodynamic conditions. Two-body potentials based on the Buckingham functional form recently developed by Wang et al. [25] and by Deng et al. [26] are two promising candidates.

By adding interaction parameters for boron to the Guillot and Sator potential[27], the Wang's potential shows a good agreement with experiments on structure and properties of borosilicate glasses and liquids. Furthermore, it has a high transferability with no composition-dependent parameters. Started from the pairwise potential developed by Kieu et al. [28], the Deng's potential well reproduced the structure and elastic moduli of borosilicate and boroaluminosilicate. However, similar to the Kieu potential, it has a composition dependent boron energy parameter that was fitted to the experimental boron coordination trends as a function of compositions. This makes it hard to transfer to the systems other than borosilicate glasses. Moreover, a benchmark work carried out recently by Lee et al. indicated that both the Wang's and Deng's potentials cannot accurately predict the elastic moduli of commercial borosilicate glasses Boro33 and N-BK7 due to the imprecise description of $N_4$ as a function of compositions [29].

Based on the Buckingham functional form, we recently used an optimization scheme similar to the one developed for silica glass [30] and extended the interaction parameter set to include alkali modifiers (Li, Na and K) and alkaline earth modifier (Ca), network former boron, and aluminum that can behave as a modifier or a former depending on the composition [31, 32]. In this work, we adapted a similar optimization approach to develop interaction parameters for Mg-O, Mg-Si, Mg-Mg, Mg-B, and Ca-B pairs. One of the major goals of our potential optimization scheme is hence to not have any system specific parameters to ensure easy transferability and extensibility to complex multi-component systems. Reliable pairwise potentials will allow for high computational efficiency to study large and complex systems.

The article is organized as follows: first, the reliability of the potential for alkaline earth silicate will be demonstrated by comparing with the *ab initio* and experimental data. Second, the structure and properties of magnesium silicate will be discussed and compared with those of calcium silicate. Afterward, we will investigate the structure and properties of magnesium and calcium borates, and then compare them with those of silicates.

## 2. Simulations Methods

In this section, we present the optimization procedure and the details about the generation of the glass samples for the investigation of composition-structure-properties relationship in alkaline earth silicates and borates.

### 2.1 Potential and cost function

As for our previous studies [30-32], we used the Buckingham potential functional form for the short-range interactions and the Wolf truncation method to evaluate the Coulombic interactions [33-35].

$$V^{Buck}(r_{\alpha\beta}) = A_{\alpha\beta} exp(-B_{\alpha\beta} r_{\alpha\beta}) - \frac{C_{\alpha\beta}}{r_{\alpha\beta}^6} + \frac{D_{\alpha\beta}}{r_{\alpha\beta}^{24}} + V^W(r_{\alpha\beta}) \quad (1)$$

where

$$V^W(r_{\alpha\beta}) = q_\alpha q_\beta \left[\frac{1}{r_{\alpha\beta}} - \frac{1}{r_{cut}^W} + \frac{(r_{\alpha\beta} - r_{cut}^W)}{(r_{cut}^W)^2}\right] \quad (2)$$

and $\alpha, \beta \in \{O, Si, B, Mg, Ca\}$. All the parameters from our previous studies [30-32] were maintained constant and oxygen charge ($q_O$) was evaluated for each composition in order to maintain the charge neutrality [36]. For instance:

$$q_O = \frac{xq_{Mg} + 2(1-x)q_B}{2x-3} \quad (3)$$

which is for magnesium borate with the composition of xMgO–(1-x)B$_2$O$_3$, where $q_{Mg}$ and $q_B$ are the charge of magnesium and boron, respectively.

The short-range interactions were cut off at 8 Å while the Coulombic interactions were cut off at 10 Å for the Wolf method, same as those used in our previous studies [30-32]. MD simulations were carried out using the LAMMPS [37] (Large-scale Atomic/Molecular Massively Parallel Simulator) with a timestep of 1.6 fs. A smaller timestep of 0.8 fs was used during the optimization process when exploring the parameter space to avoid large temporary forces that might arise.

The cost function for optimizing the parameters follows our scheme in previous work [30-32] and is given by

$$\chi^2(\phi) = w_1 \int_0^{r_{N_{RDF}}} \sum_{\alpha,\beta} (rg_{\alpha\beta}^{calc,T}(r|\phi) - rg_{\alpha\beta}^{ref,T}(r))^2 dr +$$

$$w_2(E^{calc,300K}(\phi) - E^{ref,300K})^2 + w_3(\rho^{calc,300K}(\phi) - \rho^{ref,300K})^2 +$$

$$w_4(C^{calc,300K}(\phi) - C^{ref,300K})^2 \quad (4)$$

where $\phi$ is the current parameter set, $\alpha, \beta$ are the different species, $w_1$, $w_2$, $w_3$, $w_4$ are the weights for each contribution, $rg_{\alpha\beta}(r)$ is the radial distribution function (RDF) weighted by the distance $r$ up to a maximum distance of $r_{N_{RDF}} = 7$ Å at temperature $T$ = 3500 K and 3000 K for silicate and borate, respectively. $E$ is the Young's modulus, $\rho$ is the density and $C$ is the average boron coordination at 300 K and ambient pressure, which is only included in the optimization of potential parameters for borates. The superscript "ref" refers to the first principles or experimental reference data towards which the optimization was carried out, and the superscript "calc" refers to the calculated properties using the current parameter set. The composition of 0.5MgO–0.5SiO$_2$, 0.4CaO–0.6SiO$_2$, 0.5MgO–0.5B$_2$O$_3$, and 0.5CaO–0.5B$_2$O$_3$ were used for each glass system in the potential optimization process.

The RDFs for "calc" were calculated by equilibrating a sample of 1200 atoms and 1500 atoms for magnesium silicate and calcium silicate at 3500 K, respectively, 1400 atoms for magnesium and calcium borate at 3000 K with the density given in Table I.

Samples were first equilibrated for 30 ps in the canonical (NVT) ensemble, and configurations from the next 40 ps were used to calculate RDFs.

Density and average boron coordination at room temperature were measured during the optimization process by relaxing the quenched samples of ~10000 atoms in the NPT ensemble at 300 K and ambient pressure with the current parameter set. The Young's modulus was then measured by compressing and expanding the samples at 300 K along one direction at a constant strain rate (1.25 ns$^{-1}$) up to a linear change of 0.6% and measuring the stress response:

$$E_x = \frac{d\sigma_x}{d\varepsilon_x} \quad (5)$$

where $E_x$, $\sigma_x$ and $\varepsilon_x$ are the Young's modulus, stress, and strain, respectively, along the $x$-direction.

Minimization of the cost function was performed using the Levenberg-Marquardt algorithm [38, 39] and the numerical derivatives were calculated using a finite difference method. More detail about the optimization scheme can be found in our previous studies [30-32].

Partial charges and short-range interaction parameters used in this study, including the newly optimized charge of Mg, interaction parameters of Mg-O, Mg-Si, Mg-Mg, Mg-B, and Ca-B pairs, are given in Table II and Table III, respectively. The new potentials will be referred to as "SHIK" (Sundararaman, Huang, Ispas, Kob) in the rest of the paper, following our previous studies [30-32].

**Table I:** Number of atoms and density used to equilibrate liquid at high temperatures in *ab initio* MD simulations.

| System | N (atoms) | ρ (g/cm³) |
|---|---|---|
| 0.5MgO–0.5SiO$_2$ | 400 | 2.75 |
| 0.4CaO–0.6SiO$_2$ | 390 | 2.78 |
| 0.5MgO–0.5B$_2$O$_3$ | 392 | 2.36 |
| 0.5CaO–0.5B$_2$O$_3$ | 392 | 2.50 |

**Table II:** Charge for different species.

| Species | Si | B | Ca | Mg |
|---|---|---|---|---|
| Charge (e) | 1.7755 | 1.6126 | 1.4977 | 1.085 |

**Table III:** Short-range interaction parameters.

| i-j | A$_{ij}$ (eV) | B$_{ij}$ (Å$^{-1}$) | C$_{ij}$ (eV·Å$^6$) | D$_{ij}$ (eV·Å$^{24}$) |
|---|---|---|---|---|
| O-O | 1120.5 | 2.8927 | 26.132 | 16800 |
| O-Si | 23108 | 5.0979 | 139.70 | 66.0 |
| Si-Si | 2798.0 | 4.4073 | 0.0 | 3423204 |
| O-Mg | 139373 | 6.0395 | 79.562 | 16800 |
| Si-Mg | 516227 | 5.3958 | 0.0 | 16800 |
| Mg-Mg | 19669 | 4.0 | 0.0 | 16800 |
| O-Ca | 146905 | 5.6094 | 45.073 | 16800 |
| Si-Ca | 77366 | 5.0770 | 0.0 | 16800 |
| Ca-Ca | 21633 | 3.2562 | 0.0 | 16800 |
| O-B | 16182 | 5.6069 | 59.203 | 32.0 |
| B-B | 1805.5 | 3.8228 | 69.174 | 6000.0 |
| Mg-B | 5000.0 | 4.0533 | 0.736 | 16800 |
| Ca-B | 848.55 | 5.9826 | 81.355 | 16800 |

## 2.2 Generation of *ab initio* reference data

The Vienna ab initio package (VASP) was used to perform the *ab initio* MD simulations [40, 41]. The Kohn–Sham (KS) formulation of the density functional theory with generalized gradient approximation (GGA) and the PBEsol (modified Perdew-Burke-Ernzerhof) functional was used to describe the electronic structure [42-44]. The projector-augmented-wave formalism was used for the electron-ion interaction for Kohn-sham orbitals expanded in the plane wave basis set at the Γ point of the supercell with energies up to 600 eV [45, 46]. The electronic convergence criterion for the residual minimization method-direct inversion in iterative space was fixed at $5\times10^{-7}$ eV. These parameters were chosen based on previous studies performed on various silicate and borosilicate melts and glasses [30-32, 47, 48].

*Ab initio* MD simulations were carried out in the NVT ensemble at 3500 K and 3000 K for silicates and borates, respectively, using the Nosé thermostat to control the temperature and starting from configurations obtained from equilibrium classical MD simulations at the same temperature [49]. A cubic system of N atoms with periodic boundary conditions was used with the simulation box length fixed to a value corresponding to an experimental density under ambient conditions for each composition (see details in Table I) [50-54]. For borates, a density of about 10% less than experimental glass density was chosen in order to reduce the pressure for faster diffusion during the equilibration of liquid at high temperature. The simulation for a given composition was stopped once the mean squared displacement of the slowest element, i.e., silicon/boron, reached ~10 Å$^2$, which was sufficient for other species to reach the diffusive regime too. We discarded the first 1 to 2.5 ps of the trajectory in each case and used the remaining data for calculating the reference RDFs for the potential fitting and for other structural properties presented in subsection 3.1 and 3.3.

**2.3 Generation of MD simulation samples**

Glasses of various compositions, as shown in Table IV were prepared using the melt-quench method. Samples with ~10000 atoms were first equilibrated in the NVT ensemble for about 100 ps at about the experimental glass density at 3500 K and 3000 K for silicate and borate, respectively, and then in the NPT ensemble for about 500 ps at 0.1 GPa. They were then subsequently quenched to 300 K in the NPT ensemble at a nominal quench rate of ~1 K/ps. The small pressure of 0.1 GPa was applied at high temperature as a precaution to present the system from entering the gas phase, which was ramped down to 0 GPa during the quenching process. The samples were then annealed at 300 K and 0 GPa for 100 ps in the NPT ensemble. Four independent samples were quenched for each composition to improve the statistics of the results. To investigate the pressure effect on glass, the sample was compressed/decompressed at a rate of 0.2 GPa/ps in the NPT ensemble. After every compression/decompression step, the sample was equilibrated for 90 ps, followed by another 10 ps to calculate the density.

**Table IV:** Compositions of glass systems studied in this work.

| Composition |
| --- |
| xMO–(1-x)SiO$_2$ |
| ($M \in Mg; x \in 0, 0.1, 0.2, 0.3, 0.4, 0.5, 0.66$) |
| ($M \in Ca; x \in 0, 0.1, 0.2, 0.3, 0.4, 0.5$) |
| xMO–(1-x)B$_2$O$_3$ |
| ($M \in Mg, Ca; x \in 0, 0.1, 0.2, 0.3, 0.4, 0.5$) |

## 3. Results and Discussion

In this section, the reliability of the interaction potential for magnesium and calcium silicates will be shown in subsection 3.1. The structure and properties of silicates will be discussed in the following subsection 3.2. Afterward, the reliability of the interaction potential for magnesium and calcium borates, and the structure and properties of borates will be discussed in subsection 3.3 and 3.4, respectively.

**3.1 Structure of magnesium and calcium silicate liquids**

Figure 1 shows partial RDFs of 0.5MgO–0.5SiO$_2$ and 0.4CaO–0.6SiO$_2$ liquids at 3500 K from classical MD by using the newly developed potential, in comparison with *ab initio* simulation results. The RDF data for calcium composition are extracted from our previous study [31]. Overall, the new potential well reproduces the structure of the melt as predicted by the *ab initio* simulations. This is not that surprising as RDFs were included in the cost function that was minimized. Meanwhile, it is important to note that the discrepancies observed in some of the RDFs are not entirely due to the shortcomings of the pair-wise potential functional form, as compromises in the optimization have to be made to predict different properties over a wide range of compositions [30, 31].

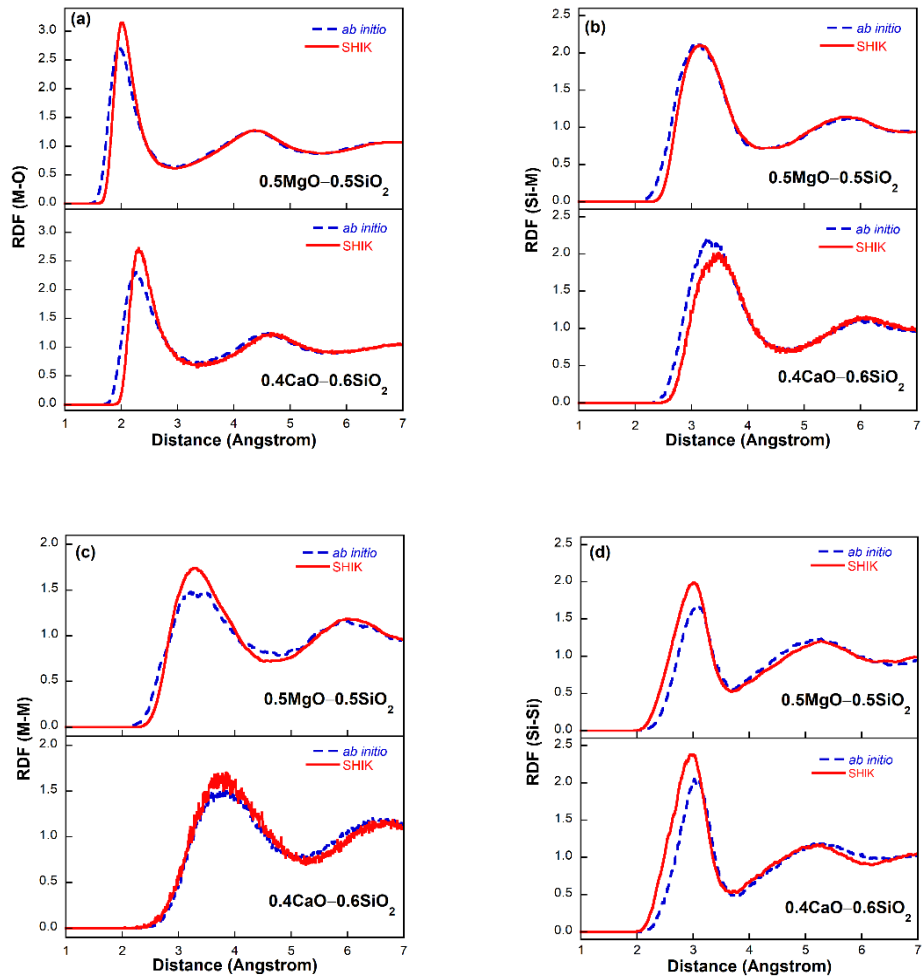

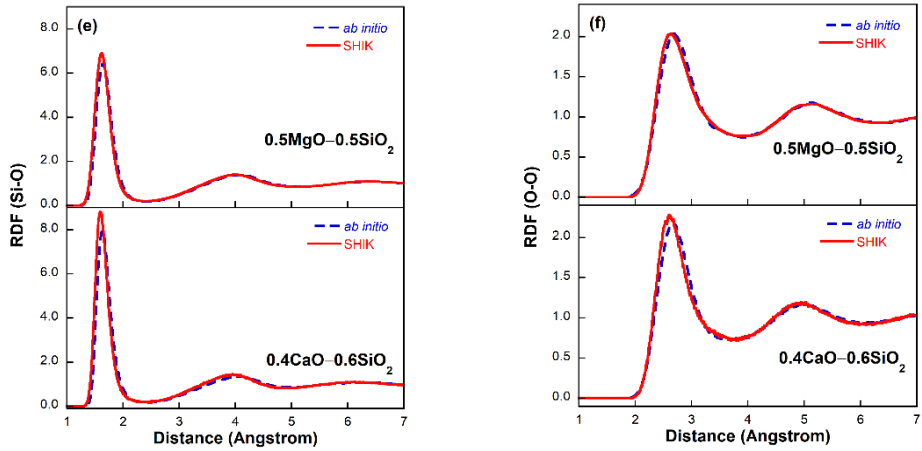

**Fig. 1** Radial distribution functions (RDFs) of (a) M-O, (b) Si-M, (c) M-M (M = Mg or Ca), (d) Si-Si, (e) Si-O, and (f) O-O pair in $0.5MgO–0.5SiO_2$ and $0.4CaO–0.6SiO_2$ liquids obtained from the *ab initio* (blue dashed line) and the SHIK potential (red solid line) simulations at 3500 K.

Figure 2 shows bond angle distributions (BADs) for the same two liquids as in Fig. 1. Although these BADs were not included in the cost function, overall a good agreement is seen between MD and *ab initio* data. The BAD of Si-O-Si and Si-Si-Si shift slightly to the left in both magnesium silicate and calcium silicate liquids, consistent with the shorter Si-Si distance in classical MD as seen in Fig.1(d). This indicates a more rigid glass network structure in classical MD compared to that in *ab initio* simulation. Moreover, the BAD of Si-Si-Si exhibits a higher peak at 60 degrees, suggesting more 3-membered rings in the classical MD.

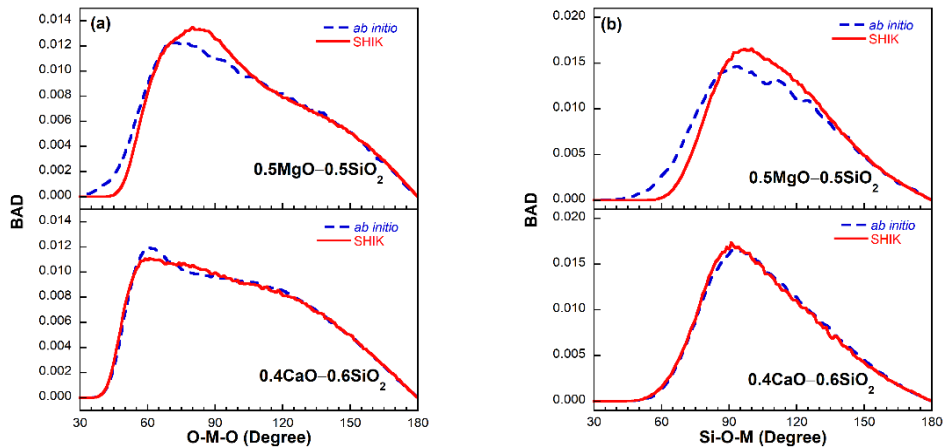

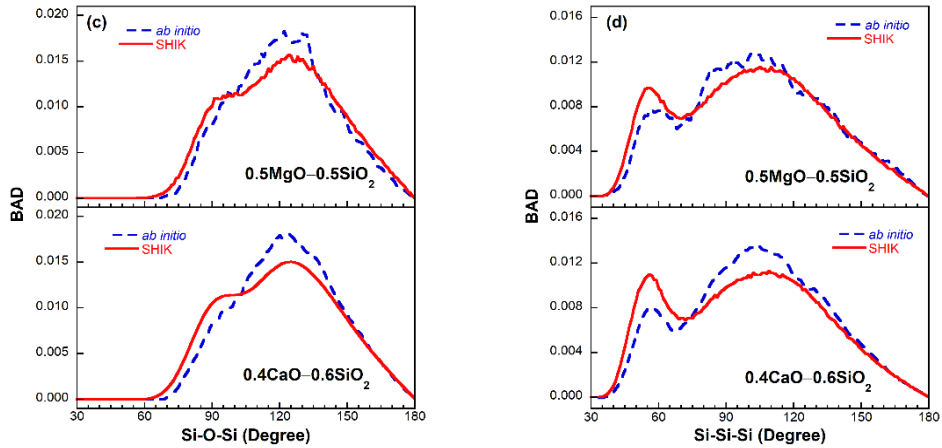

**Fig. 2** (a) O-M-O, (b) Si-O-M (M= Mg or Ca), (c) Si-O-Si, and (d) Si-Si-Si bond angle distributions (BADs) in 0.5MgO–0.5SiO$_2$ and 0.4CaO–0.6SiO$_2$ liquids obtained from the *ab initio* (blue dashed line) and the SHIK potential (red solid line) simulations at 3500 K.

### 3.2 Structure and properties of magnesium and calcium silicate glasses

In this subsection, we will present a number of structural signatures and mechanical properties of magnesium and calcium silicates calculated from MD simulations and compare with existing experimental data to show the reliability of the new potential to describe them over a large composition range and under different thermodynamic conditions.

The Q$^n$ species indicate the degree of polymerization of the silica network, where *n* is the number of bridging oxygen (BO) in the SiO$_4$ tetrahedron. As seen from the Q$^n$ distribution in Fig. 3(a), the polymerization of the glass network is decreased with increasing modifier content. Moreover, magnesium silicate (solid symbol) exhibits a slightly higher degree of polymerization/connectivity than calcium silicate (open symbol) at a given modifier content.

A similar trend is observed in the fraction of different oxygen species in Fig. 3(b), the BO decreases and NBO increases with the increasing modifier content. Magnesium silicate has a slightly higher fraction of BO and a lower fraction of NBO than calcium silicate at a given modifier content, indicating a slightly higher degree of the polymerization in the glass network in the former than in the latter. In addition, free oxygen (FO), the oxygen not bonded to any silicon, increases with the increasing modifier content and has a slightly higher fraction in magnesium silicate than in the calcium silicate at higher modifier content. This observation indicates that the high field strength of Mg can better stabilize the local negative charge and thus promote the formation of FOs, suggesting that Mg ions are incorporated into the glass network.

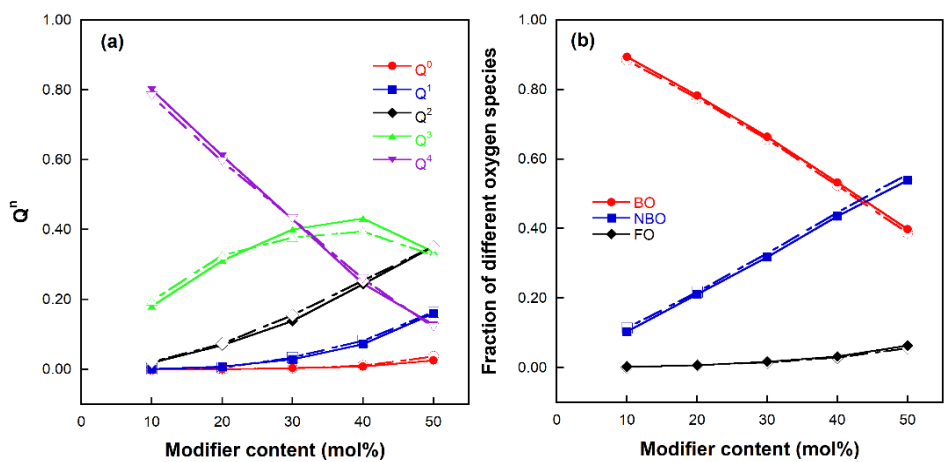

**Fig. 3** (a) $Q^n$ species and (b) fraction of different oxygen species as a function of modifier content in magnesium (solid symbol) and calcium (open symbol) silicate glasses at 300 K from classical MD. Error bars are smaller than the symbols.

The average bond length inside $SiO_4$ tetrahedron as a function of modifier content in magnesium silicate and calcium silicate are shown in Fig. 4. At a given modifier content, calcium silicate exhibits shorter O-O, Si-BO, and Si-NBO bond length than magnesium silicate. The O-O bond length decreases with increasing modifier content in magnesium silicate and calcium silicate; whereas, magnesium silicate exhibits a smaller decreasing trend than calcium silicate. Similar to the O-O bond length, both the Si-BO and Si-NBO bond length decrease obviously with increasing modifier content in calcium silicate and exhibit insignificant change with increasing modifier content in magnesium silicate. This indicates that $SiO_4$ tetrahedra are less perturbed by the addition of modifiers in magnesium silicate than in calcium silicate.

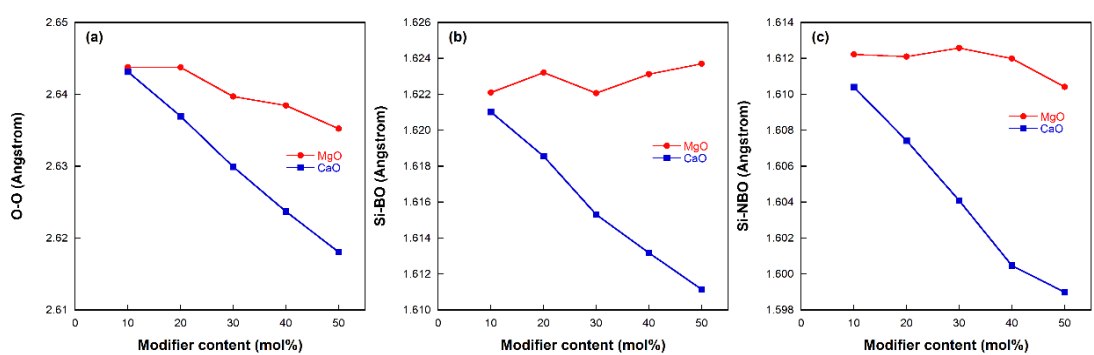

**Fig. 4** Average (a) O-O, (b) Si-BO, and (c) Si-NBO bond length as a function of modifier content in magnesium and calcium silicate glasses at 300 K from classical MD. Error bars are smaller than the symbols.

Figure 5 shows the inter-tetrahedral bond length and bond angle between SiO$_4$ tetrahedra, and the structural information between the modifier and SiO$_4$ tetrahedron. Larger inter-tetrahedral (Si-O-Si) angle, O-M-O (M = Mg or Ca) and Si-O-M angle are observed in magnesium silicate than those in calcium silicate at a given modifier content. In contrast to the obvious decreasing trend of Si-O-Si and Si-O-M, the O-M-O exhibits a very small change with increasing modifier content. At a given modifier content, the Si-Si bond length in magnesium silicate is longer than that in calcium silicate, whereas the opposite is true for Si-M and M-O bond length. In addition, the Si-M and Si-Si bond length change significantly, while the M-O bond length changes very little with increasing modifier content. The less obvious decreasing trend in the Si-O-M bond angle and the shorter Si-M bond length in magnesium silicate (Fig. 5(c) and (d)) suggest a more rigid connection between SiO$_4$ tetrahedra and modifier ions.

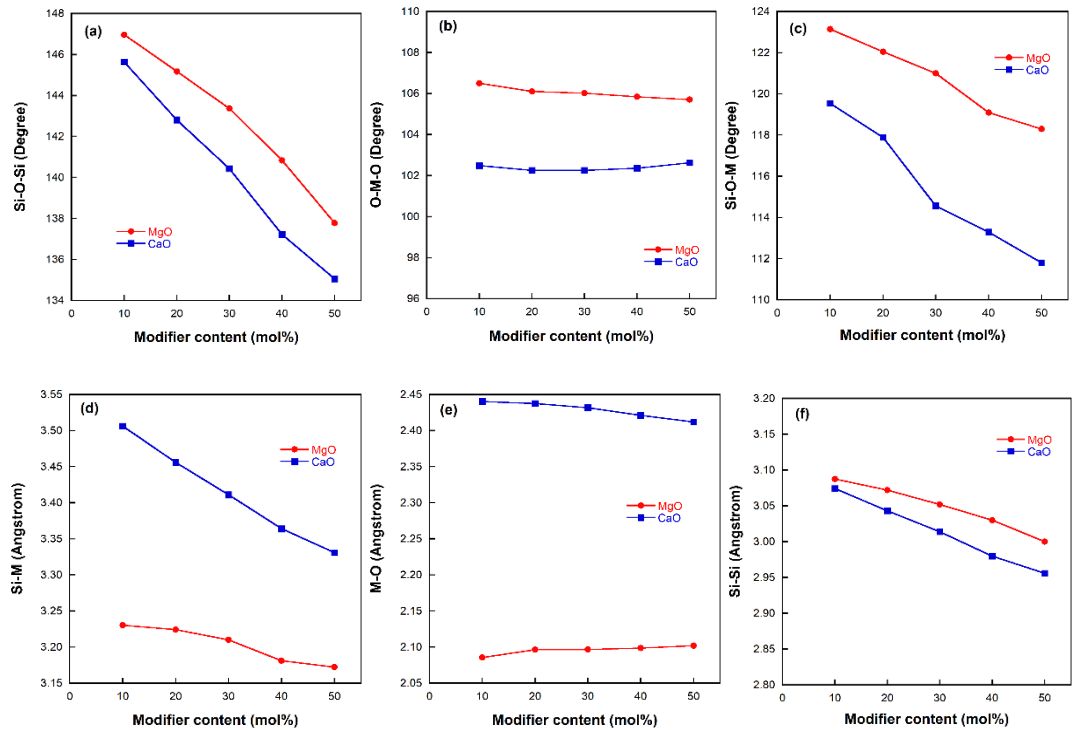

**Fig. 5** Average (a) Si-O-Si, (b) O-M-O (M = Mg or Ca), and (c) Si-O-M bond angle; (d) Si-M, (e) M-O, and (f) Si-Si bond length as a function of modifier content in magnesium and calcium silicate glasses at 300 K from classical MD. Error bars are smaller than the symbols.

In Fig. 6, the coordination number of Mg and Ca in magnesium silicate and calcium silicate are in the range of 4.4~5 and 5.3~5.9, respectively, consistent with previous studies [22, 55-57]. The coordination number of Mg in magnesium silicate increases with increasing modifier content, whereas the coordination number of Ca in calcium silicate exhibits an increasing trend in the composition range of 10~40 mol% CaO, and then decreases afterwards. The lower coordination number of magnesium manifests its covalent character inside the glass network, which suggests a stronger bond strength between the $SiO_4$ tetrahedron and the modifier.

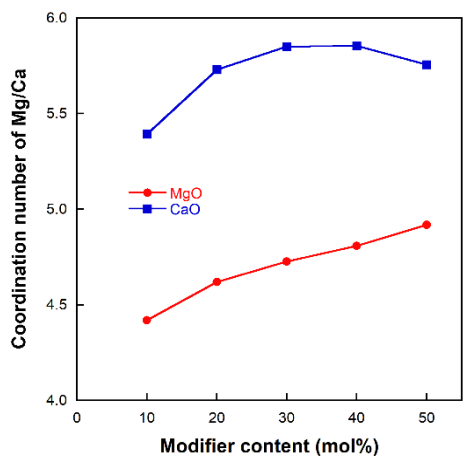

**Fig. 6** Coordination number of Mg and Ca in magnesium and calcium silicate glasses as a function of modifier content at 300 K from classical MD.

Figure 7 shows the primitive ring statistics in magnesium silicate and calcium silicate calculated by using the R.I.N.G.S. program [58]. Here, the primitive rings are defined as the shortest closed loop that includes a given Si atom and two of its nearest neighbor O atoms in the silica network [59, 60]. In comparison with the ring statistics in silica glass, the ring size distribution shifts toward small-membered rings with increasing modifier content in both magnesium and calcium silicate. The rather high 3- and 4- membered rings in $0.5CaO–0.5SiO_2$ glass are attributed to the peak at 60 and 90 degrees in the Si-Si-Si BAD, respectively [61] as shown in Fig. 7(d), in comparison with that in magnesium silicate (Fig. 7(c)). Meanwhile, magnesium silicate has more large-sized rings at a given modifier content, and shows less modification to the ring statistics in silica glass at a given modifier content, in a good agreement with the observations in the previous reverse Monte Carlo study [18]. This suggests that Mg ions are incorporated into the glass network, hence increasing the connectivity of the overall glass network [62-64].

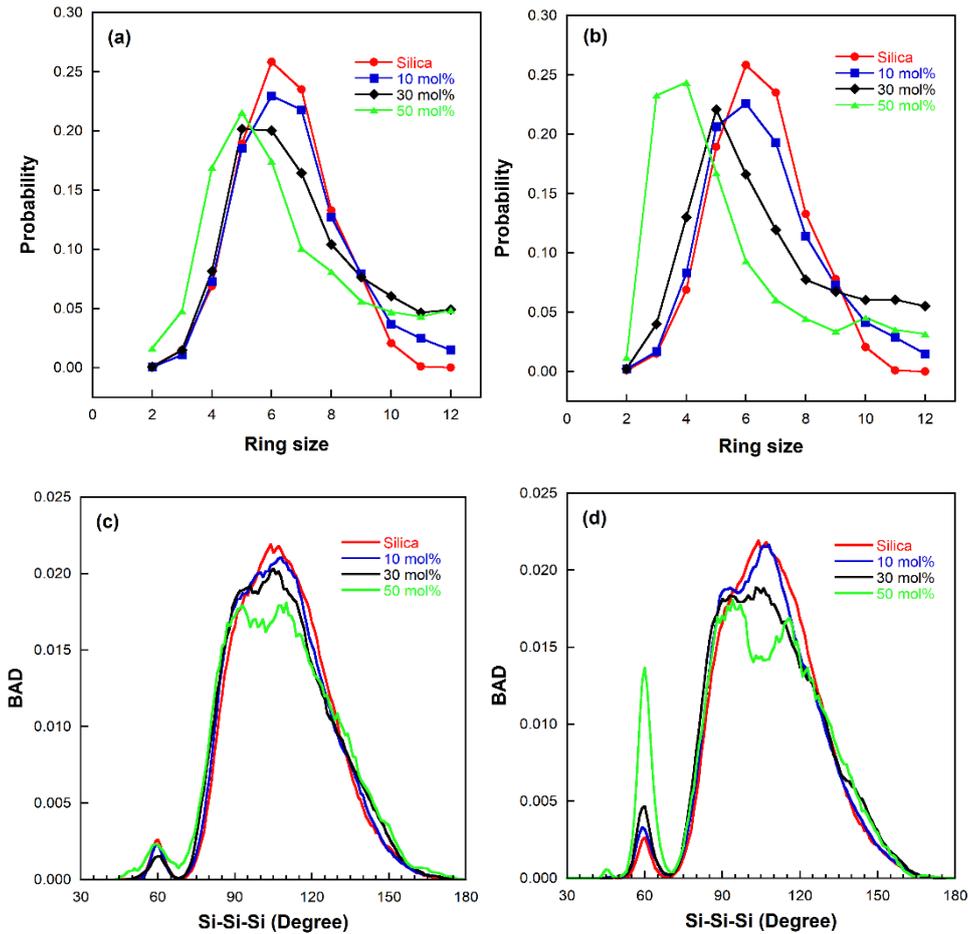

**Fig. 7** Ring statistic of (a) magnesium silicate and (b) calcium silicate, Si-Si-Si BAD of (c) magnesium silicate and (d) calcium silicate as a function of modifier content at 300 K from classical MD, in comparison with that in silica glass.

Density and elastic moduli from experiments and MD simulations are plotted in Fig. 8. It should be noted, while there are some investigations on calcium silicate glasses in experiments [50, 65-72], studies on magnesium silicate glasses are very limited [15], mostly focused on MgO–SiO$_2$ (MgSiO$_3$, enstatite) and 2MgO–SiO$_2$ (Mg$_2$SiO$_4$, forsterite) due to their geological importance. Figure 8 shows that both density and elastic moduli increase with increasing modifier content. At a given modifier content, the density of magnesium silicate is slightly lower than that of calcium silicate except MgO–SiO$_2$, whereas the elastic moduli of magnesium silicate are higher than those of calcium silicate in the composition range investigated here. In general, a good agreement is seen between experiments [15, 50, 65-72] and MD simulations, although some difference (<10%) in elastic moduli can be seen in Fig. 8 (b) and (c).

In short, the competition between the depolymerization of the glass network and the formation of new bonds with the modifier leads to the enhancement of the elastic moduli with increasing modifier content in alkaline earth silicate glasses. The higher elastic moduli of magnesium silicate compared to calcium silicate may result from the higher connectivity of the overall glass network due to the incorporation of fourfold coordinated magnesium and a more rigid connection between $SiO_4$ tetrahedra and modifier ions.

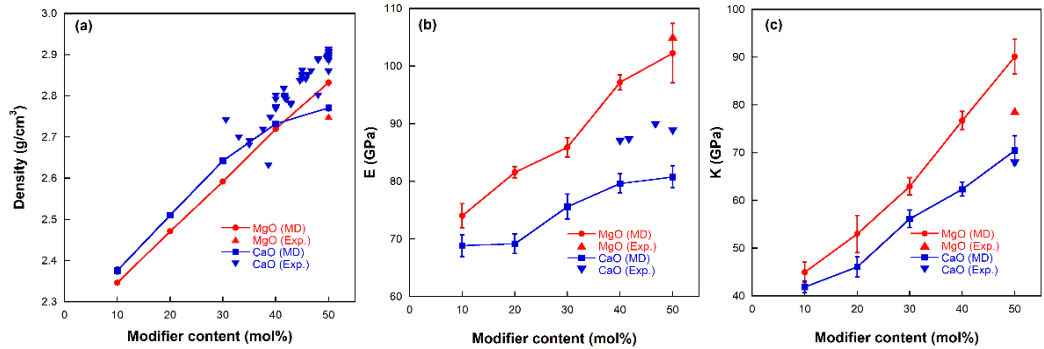

**Fig. 8** (a) Density, (b) Young's modulus ($E$), and (c) bulk modulus ($K$) as a function of modifier content in magnesium and calcium silicate glasses at ambient conditions from experiments [15, 50, 65-72] and classical MD.

To further validate the reliability of the newly parameterized interaction potential under different thermodynamic conditions, the structure and properties of $MgO$–$SiO_2$ (enstatite), $CaO$–$SiO_2$ (wollastonite), and $2MgO$–$SiO_2$ (forsterite) glasses were investigated and compared with available experimental results [73, 74]. The structure factor ($S(q)$) of vitreous enstatite, wollastonite, and forsterite under different pressures are shown in Fig. 9. Overall, the structure factors from MD simulation well reproduce the experimental neutron structure factors of vitreous enstatite and wollastonite, and X-ray structure factors of vitreous forsterite [73, 74]. In all three glasses, the first sharp diffraction peak (FSDP) shifts to a higher $q$ value with increasing pressure. In vitreous enstatite and wollastonite, the intensity of the principal peak around 3 Å$^{-1}$ increases with increasing pressure. For vitreous forsterite, the FSPD sharpens under pressure and an additional peak around 3 Å$^{-1}$ appears under 20 GPa, which has been associated with the formation of $SiO_6$ octahedra [75].

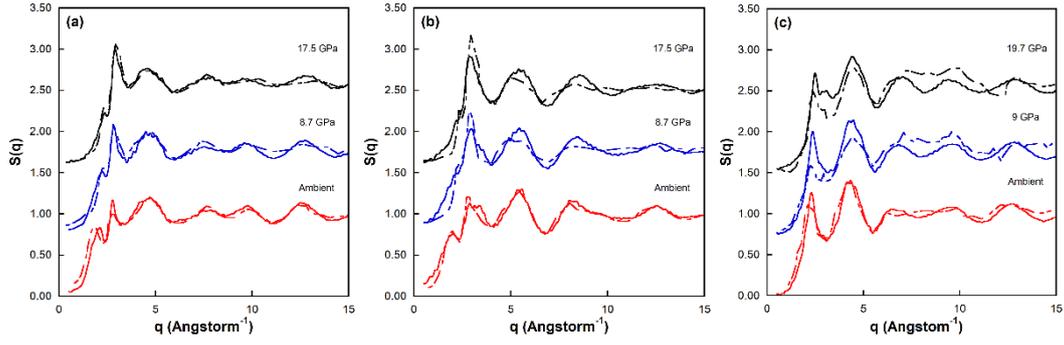

**Fig. 9** Neutron structure factor (S(q)) of (a) MgO–SiO$_2$ (enstatite), (b) CaO–SiO$_2$ (wollastonite), and X-ray structure factor of (c) 2MgO–SiO$_2$ (forsterite) glasses under different pressures at room temperature from classical MD (solid line) in comparison with experiments [73, 74] (dashed line). Structure factors at high pressures are shifted vertically for clarity.

Figure 10 shows that density and elastic moduli of MgO–SiO$_2$ glass increase with increasing pressure in classical MD simulations, consistent with Brillouin light scattering experiments [69]. The larger difference in density between simulation and experiment in the high pressure region (>10 GPa) may be attributed to the onset of irreversible densification that leads to the underestimated density calculated from sound velocities measured in Brillouin light scattering experiments, which in turn gives lower elastic moduli. This is confirmed by the compression-decompression curve in Fig. 11 (a). The irreversible densification becomes more obvious around 8~10 GPa, corresponding to the pressure at which the population of five-coordinated silicon starts to increase rapidly with pressure as seen in the Fig. 11 (b).

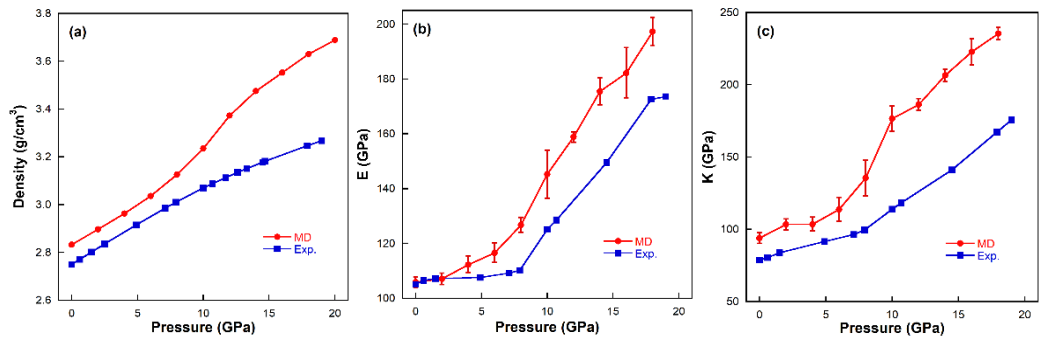

**Fig. 10** Variation of (a) density, (b) Young's modulus (*E*), and (c) bulk modulus (*K*) of MgO–SiO$_2$ glass with pressure at room temperature from classical MD in comparison with experiments [69].

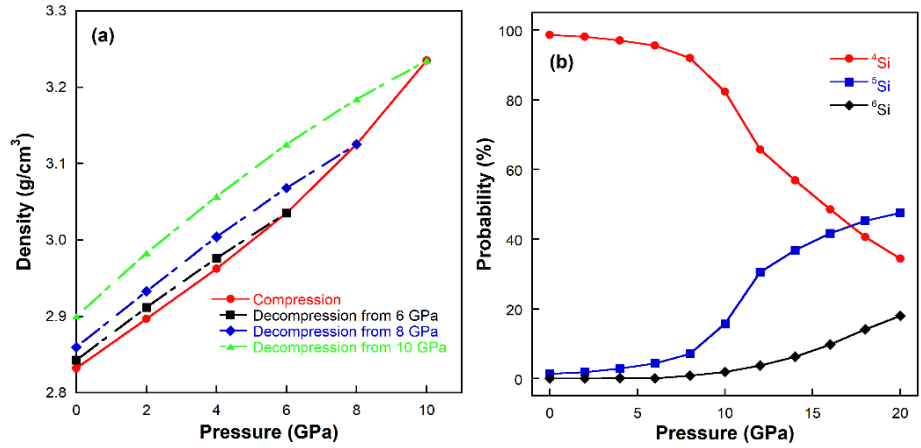

**Fig. 11** (a) Variation of density of MgO–SiO$_2$ glass with pressure during compression-decompression, (b) variation of coordination number of Si of MgO–SiO$_2$ glass with pressure at 300 K from classical MD.

### 3.3 Structure of magnesium and calcium borate liquids

Figure 12 shows the partial RDFs of 0.5MgO–0.5B$_2$O$_3$ and 0.5CaO–0.5B$_2$O$_3$ liquids at 3000 K from MD simulations by using the newly parameterized potential, in a good agreement with *ab initio* simulation results.

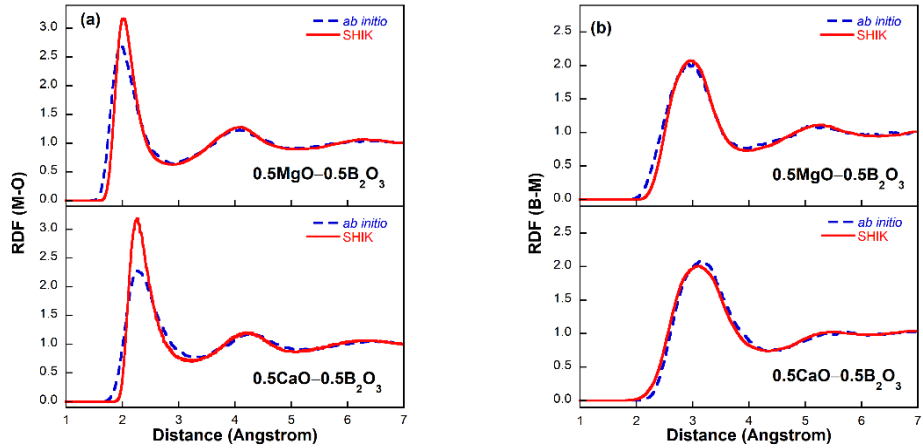

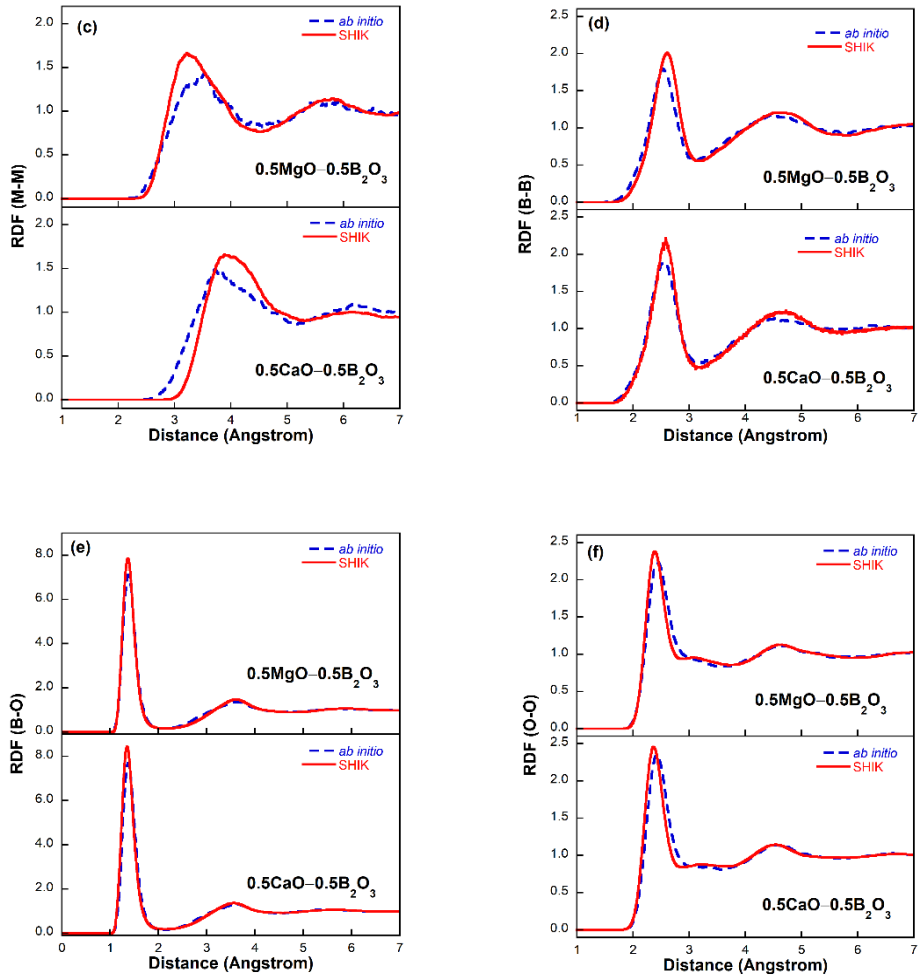

**Fig. 12** RDFs of (a) M-O, (b) B-M, (c) M-M (M = Mg or Ca), (d) B-B, (e) B-O, and (f) O-O pairs in $0.5MgO–0.5B_2O_3$ and $0.5CaO–0.5B_2O_3$ liquids obtained from the *ab initio* (blue dashed line) and the SHIK potential (red solid line) simulations at 3000 K.

Figure 13 shows the BADs in $0.5MgO–0.5B_2O_3$ and $0.5CaO–0.5B_2O_3$ liquids at 3000 K by using the newly parameterized potential, overall a good agreement is seen between classical MD and *ab initio* data. The O-M-O bond angle from the classical MD simulation exhibits a more obvious two peaks structure, which may correspond to the splitting of BO-M-BO and NBO-M-NBO bond angle [48, 76]. The BAD of B-O-B shifts to the high-angle distribution from classical MD simulation in comparison to that from *ab initio* simulation, which implies a more open structure in the former than in the latter. The same discrepancy in the BAD of B-O-B was observed in our previous work on alkali borates. As we used the same B-B parameters in the current work, this may be resulted from a deficiency from the previous optimization [32].

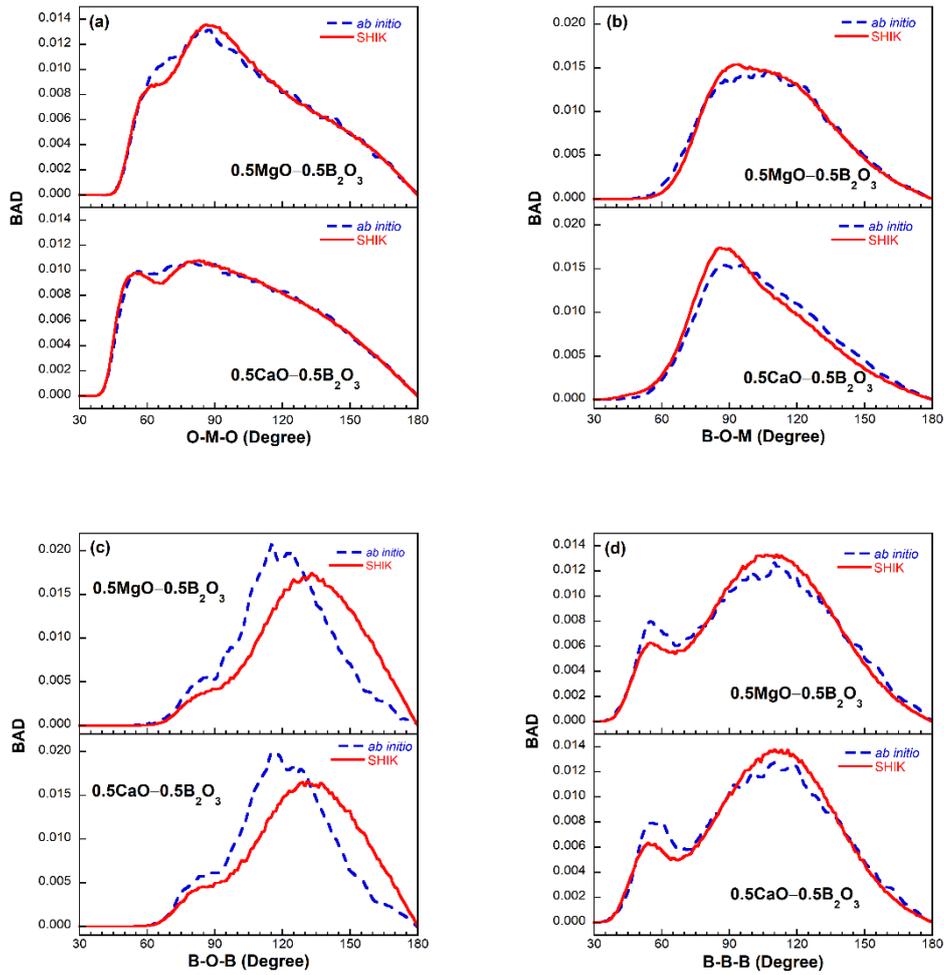

**Fig. 13** (a) O-M-O, (b) B-O-M (M= Mg or Ca), (c) B-O-B, and (d) B-B-B BADs in 0.5MgO–0.5B$_2$O$_3$ and 0.5CaO–0.5B$_2$O$_3$ liquids obtained from the *ab initio* (blue dashed line) and the SHIK potential (red solid line) simulations at 3000 K.

### 3.4 Structure and properties of magnesium and calcium borate glasses

Figure 14 shows the fraction of fourfold coordinated born ($N_4$) in magnesium and calcium borate glasses. A good agreement is seen between simulations and experiments in the composition range studied [24]. Moreover, the difference in the $N_4$ values between calcium borate and magnesium borate is well reproduced by the newly parameterized potential. In the narrow composition range studied in experiments, the $N_4$ changes very little in magnesium borate, with values close to what was predicted in classical MD simulation at 50 mol% of MgO content.

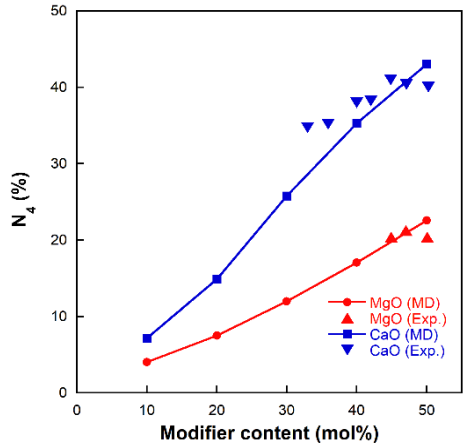

**Fig. 14** Fraction of fourfold coordinated boron ($N_4$) as a function of modifier content in magnesium and calcium borate glasses at ambient conditions from classical MD simulations and experiments [24].

Figure 15 shows the $Q^n$ species and fraction of different oxygen species in magnesium borate (solid symbol) and calcium borate (open symbol). Overall, magnesium borate shows a slightly larger fraction of boron in most of the $Q^n$ species, while the $Q^4$ of $BO_4$ in calcium borate is much higher than that in magnesium borate. The fraction of different oxygen species in Fig. 15(c) indicates the formation of more NBOs and FOs in magnesium borate compared to calcium borate.

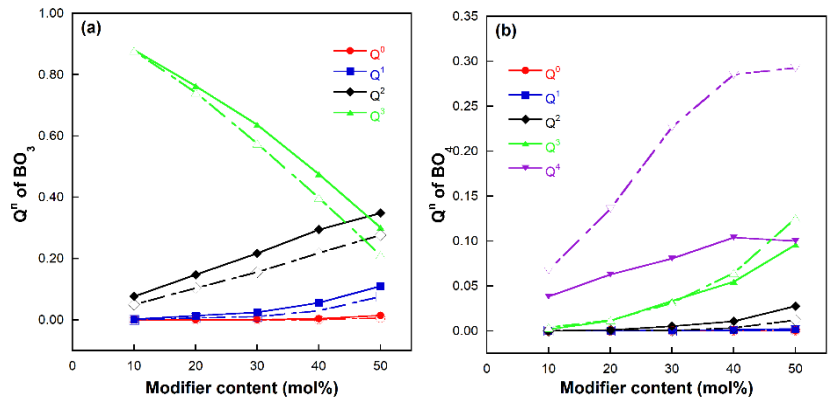

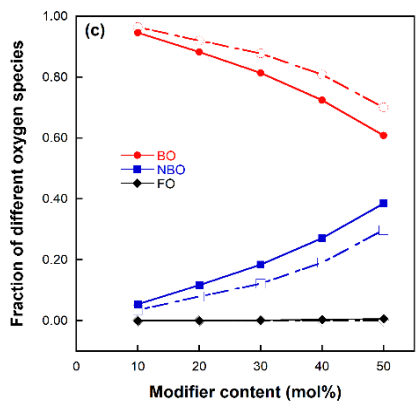

**Fig. 15** $Q^n$ species in (a) $BO_3$ and (b) $BO_4$ units, and (c) fraction of different oxygen species as a function of modifier content in magnesium (solid symbol) and calcium (open symbol) borate glasses at 300 K from classical MD.

The B-O bond length in $BO_3$ ($^3$B-O) and $BO_4$ ($^4$B-O) units in magnesium borate and calcium borate are shown in Fig. 16. Overall, the $^3$B-O bond length is shorter than that of the $^4$B-O bond, which is in good agreement with the experimental data [77]. At a given modifier content, calcium borate exhibits a shorter bond length in both $^3$B-O and $^4$B-O. Moreover, the bond length of $^3$B-O and $^4$B-O decrease with increasing modifier content and exhibit a less obvious change especially in $^3$B-O units in the magnesium borate. The coordination number of Mg and Ca in magnesium borate and calcium borate as a function of modifier content are shown in Fig. 17, which are higher than those in silicates in Fig. 6. Within the composition range of 10~50 mol% modifier content, the coordination number of Mg and Ca exhibits an insignificant change with increasing modifier content.

The higher NBOs and FOs fraction (Fig. 15 (c)) and the lower coordination number of modifier (Fig. 17) in magnesium borate suggest that the magnesium may enter the $B_2O_3$ network instead of converting $BO_3$ to $BO_4$, thus leading to much lower $N_4$ (Fig. 14) and $Q^4$ in $BO_4$ (Fig. 15 (b)) comparing to those in calcium borate with similar modifier contents. This is further verified from the insignificant decrease of B-O bond length with modifier content in magnesium borate, which may be attributed to the lower conversion rate of $BO_3$ to $BO_4$ units (Fig. 16).

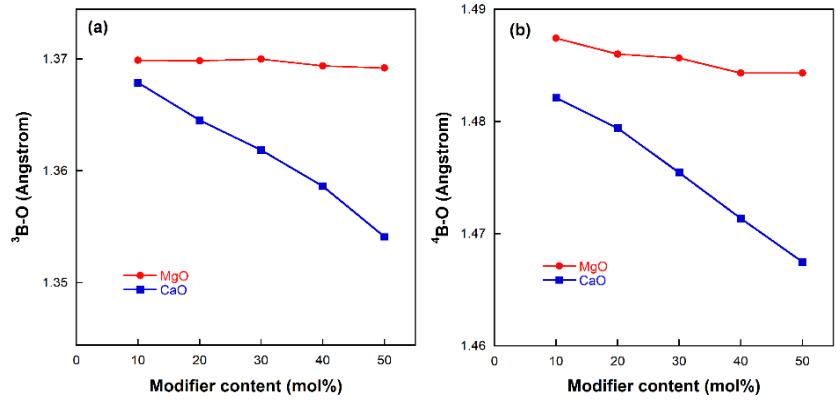

**Fig. 16** (a) $^3$B-O and (b) $^4$B-O bond length as a function of modifier content in magnesium and calcium borate glasses at 300 K from classical MD.

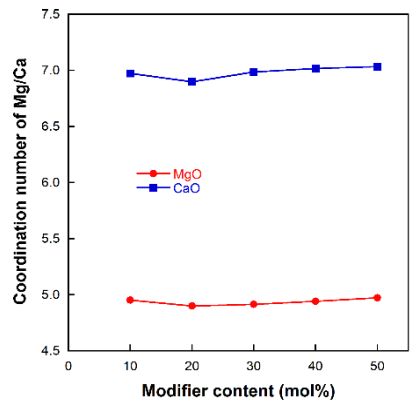

**Fig. 17** Coordination number of Mg and Ca in magnesium and calcium borate glasses as a function of modifier content at 300 K from classical MD.

Figure 18 shows that density and elastic moduli of magnesium and calcium borate glasses increase with increasing modifier content, in a good agreement with experimental results [51-54, 78-81]. Density and elastic moduli of magnesium borate are lower than those of calcium borate at a given modifier content. In contrast to alkaline earth silicates, the effect of modifier content on the elastic moduli of alkaline earth borates is dominated by the $N_4$. The larger amount of $N_4$ (Fig. 14) and the shorter $^3$B-O and $^4$B-O bond length (Fig. 16) in calcium borate result in a more rigid network structure, thus higher elastic moduli as seen in Fig. 18(b)-(c).

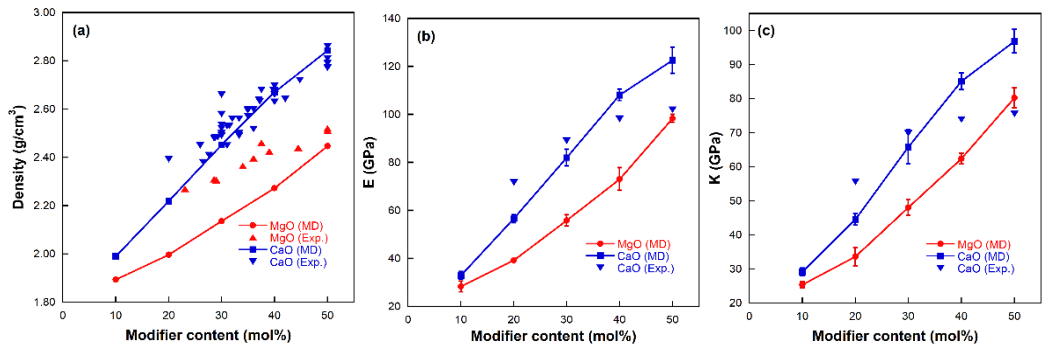

**Fig. 18** (a) Density, (b) Young's modulus ($E$), and (c) bulk modulus ($K$) as a function of modifier content in magnesium and calcium borate glasses from MD simulations at ambient conditions in comparison with experiments [51-54, 78-81].

## 4. Conclusions

The newly parameterized interatomic potentials for alkaline earth silicate and borate were used to study the structure and properties of magnesium silicate and borate over a large composition range and compared with those of calcium silicate and borate. The competition between the depolymerization of the glass network and the formation of new bonds with the modifiers leads to the enhancement of the elastic moduli with increasing modifier content in alkaline earth silicate glasses. The higher elastic moduli in magnesium silicate may result from the higher connectivity of the over glass network due to the incorporation of fourfold coordinated magnesium and a more rigid connection between $SiO_4$ tetrahedra and modifier ions. In contrast to the silicates, the effect of modifier content on the elastic moduli of alkaline earth borates is dominated by the $N_4$. Calcium borate with higher $N_4$ shows a more rigid network structure and higher elastic moduli.

## Acknowledgments

This work was supported by the National Science Foundation under grant No. DMR-1508410 and DMR-1936368. We thank Dr. Walter Kob at the University of Montpellier for his contribution to the development of the earlier version of the SHIK potential and for his continued support for this work.

## Data Availability

The data that support the findings of this study are available from the corresponding author upon reasonable request.

## Declaration of Competing Interest

The authors declare that they have no known competing financial interests or personal relationships that could have appeared to influence the work reported in this paper.

Physics, 120 (2010) 381-386.